\documentclass[11pt, a4paper]{article}
\usepackage{moriond}
\usepackage{multicol}
\usepackage{here}

\newif\ifpdf
\ifx\pdfoutput\undefined
    \pdffalse           
\else
    \pdfoutput=1        
    \pdftrue
\fi

\ifpdf
\usepackage[pdftex]{graphicx}
\else
\usepackage{graphicx}
\fi

\begin{document}
\vspace*{4cm}
\title{MEASUREMENT OF MASSES AND LIFETIMES OF B HADRONS}

\author{F.~FILTHAUT for the D\O\ and CDF Collaborations}
\address{Radboud University, Nijmegen and NIKHEF\\
Toernooiveld 1, 6525 ED Nijmegen, The Netherlands}

\maketitle
\abstracts{We present recent measurements by the CDF and D\O\ Collaborations at
  the Tevatron Collider on the masses and lifetimes of B hadrons. The results
  are compared to predictions based on Heavy Quark Effective Theory, lattice
  gauge theory, and quark models.}

\section{Introduction}
\label{sec:intro}

The ongoing Run II at the Tevatron $\mathrm{p\bar{p}}$ Collider at the Fermi
National Accelerator Laboratory has produced a wealth of results on
B-physics. While the physics of $\mathrm{B_{d}}$ and $\mathrm{B^{\pm}}$ mesons
is largely the domain of the $\mathrm{e^{+}e^{-}}$ B-factories operating at the
$\Upsilon(4S)$ resonance, access to the heavier B hadrons is presently reserved
exclusively for the Tevatron. The drawback of working in the less clean hadron
collider environment is more than compensated for by the high total
$\mathrm{b\bar{b}}$ production cross section (of order 100 $\mu$b) and the high
luminosity (the integrated luminosity delivered to the CDF and D\O\ experiments
to date is of order 2 fb$^{-1}$).

In the following, we discuss recent results obtained thanks to the high
statistics B samples. Sect.~\ref{sec:spectroscopy} discusses
new excited states that have been identified in the B, $\mathrm{B_{s}}$, and
$\Lambda_{\mathrm{b}}$ systems. Sect.~\ref{sec:lifetimes} covers a number of new
and precise lifetime measurements. The measurements have been made on data
samples of integrated luminosity ranging from 0.3 to 1.3 fb$^{-1}$. The two
experiments' results are compared, both with each other and theoretical
predictions.

\section{Spectroscopy}
\label{sec:spectroscopy}

\subsection{Orbitally excited $B$ and $B_{s}$ mesons}
\label{sec:orbitally-excited-b}

The spectroscopy of bound states containing one heavy quark ($Q$) and a light
antiquark\footnote{Here and in the following, charge conjugated states are
  implied.} is of great interest to quark models, as the dynamics of the light
antiquark becomes independent of $m_{Q}$, and the total and light quark's
angular momentum become good independent quantum numbers.
However, low statistics prevented the $L=1$ orbitally excited
$\mathrm{B^{\ast\ast}}$ states from being investigated in great detail
previously.

The $\mathrm{B^{\ast\ast}}$ system is thought to consist of four states, two of
which have a light quark angular momentum $J_{q} = 1/2$; their decay to
$\mathrm{B}^{(\ast)}\pi$ proceeds via an $S$-wave, and their total
decay width is expected to be $\mathcal{O}(100 \mbox{MeV})$, too wide to be
observed unambiguously. The $J_{q} = 3/2$ $\mathrm{B_{1}}$ and
$\mathrm{B_{2}^{\ast}}$ states (with total spin 1 and 2, respectively) decay via
a $D$-wave, and their width, of order 10 MeV, should allow them to be
identified clearly.

Both experiments have searched for the $\mathrm{B_{1}}$ and
$\mathrm{B_{2}^{\ast}}$ states, through the decays $\mathrm{B_{1}^{0}\rightarrow
B^{\ast+}\pi^{-}}$ and $\mathrm{B_{2}^{\ast 0}\rightarrow
B^{(\ast)+}\pi^{-}}$. The photon from the decay
$\mathrm{B^{\ast+}\rightarrow B^{+}\gamma}$ is not observed, resulting in two
$\mathrm{B_{2}^{\ast}}$ peaks in the B$\pi$ invariant mass spectrum, displaced
by 46 MeV. Both experiments identified $\mathrm{B^{+}}$ mesons through their
decay to $J/\psi \mathrm{K}^{+}$, the $J/\psi$ being recognized easily in its
decay to muons; in addition, CDF employed the decay $\mathrm{B^{+}\rightarrow
  \overline{D}^{0}\pi^+}$.

\begin{figure}[htb]
  \centering
  \includegraphics[width=0.42\textwidth]{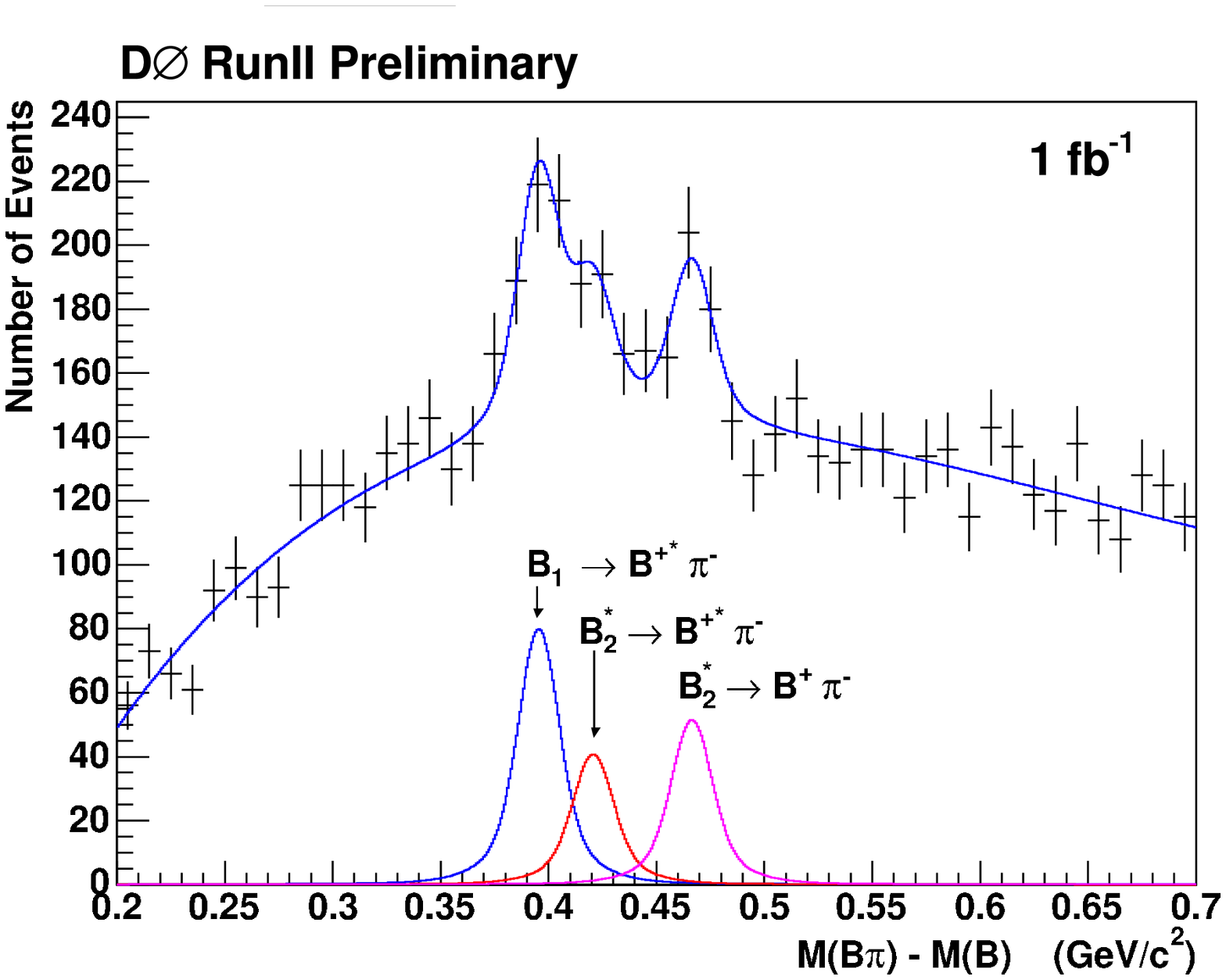}
  \includegraphics[width=0.54\textwidth]{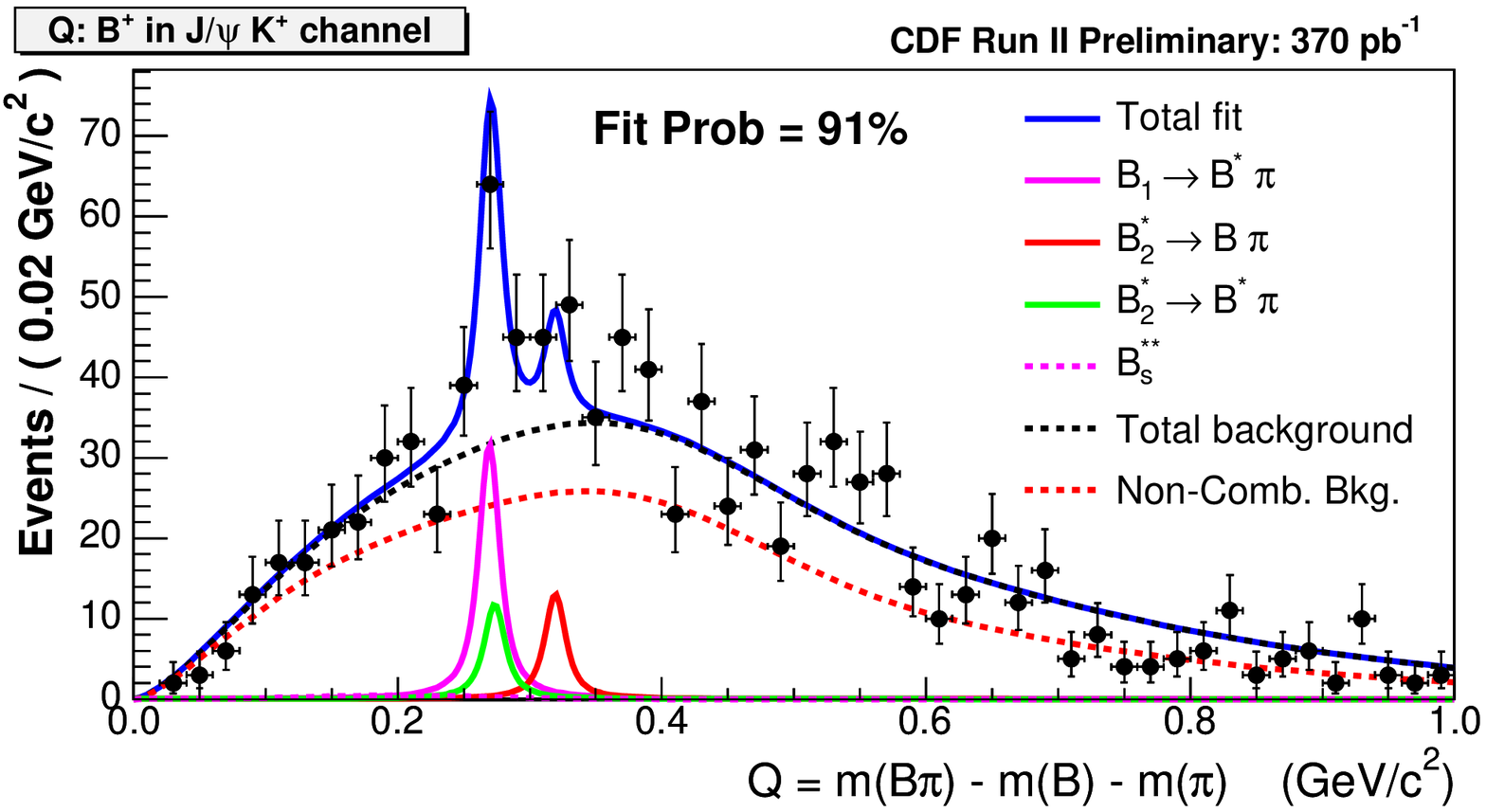}
  \caption{Distributions of the invariant mass difference $m(\mathrm{B}\pi) -
    m(\mathrm{B})$ (left) for D\O, and $m(\mathrm{B}\pi) -
    m(\mathrm{B}) - m(\pi)$ (right) for CDF, reconstructed in the mode
    $\mathrm{B^{+}}\rightarrow J/\psi \mathrm{K^{+}}$.}
  \label{fig:Bpi}
\end{figure}

The resulting $\mathrm{B\pi}$ invariant mass spectra, for the $J/\psi
\mathrm{K}^{+}$ decay mode, are shown in Fig.~\ref{fig:Bpi}. CDF's results are
consistent with those using the $\mathrm{\overline{D}}^{0}\pi^+$ mode. The two
experiments' results obviously disagree: while D\O\ find clear 
evidence for three separately observed peaks, CDF find that the
$\mathrm{B_{1}^{0},\;B_{2}^{\ast 0}\rightarrow B^{\ast+}\pi^{-}}$ peaks
coincide. The resulting $\mathrm{B_{1}^{0}}$--$\mathrm{B_{2}^{\ast 0}}$ mass
splittings (25 MeV and 4 MeV, respectively) both disagree with the theory
prediction of 12--14 MeV~\cite{eichten}.

\begin{figure}[htb]
  \centering
  \includegraphics[width=0.40\textwidth]{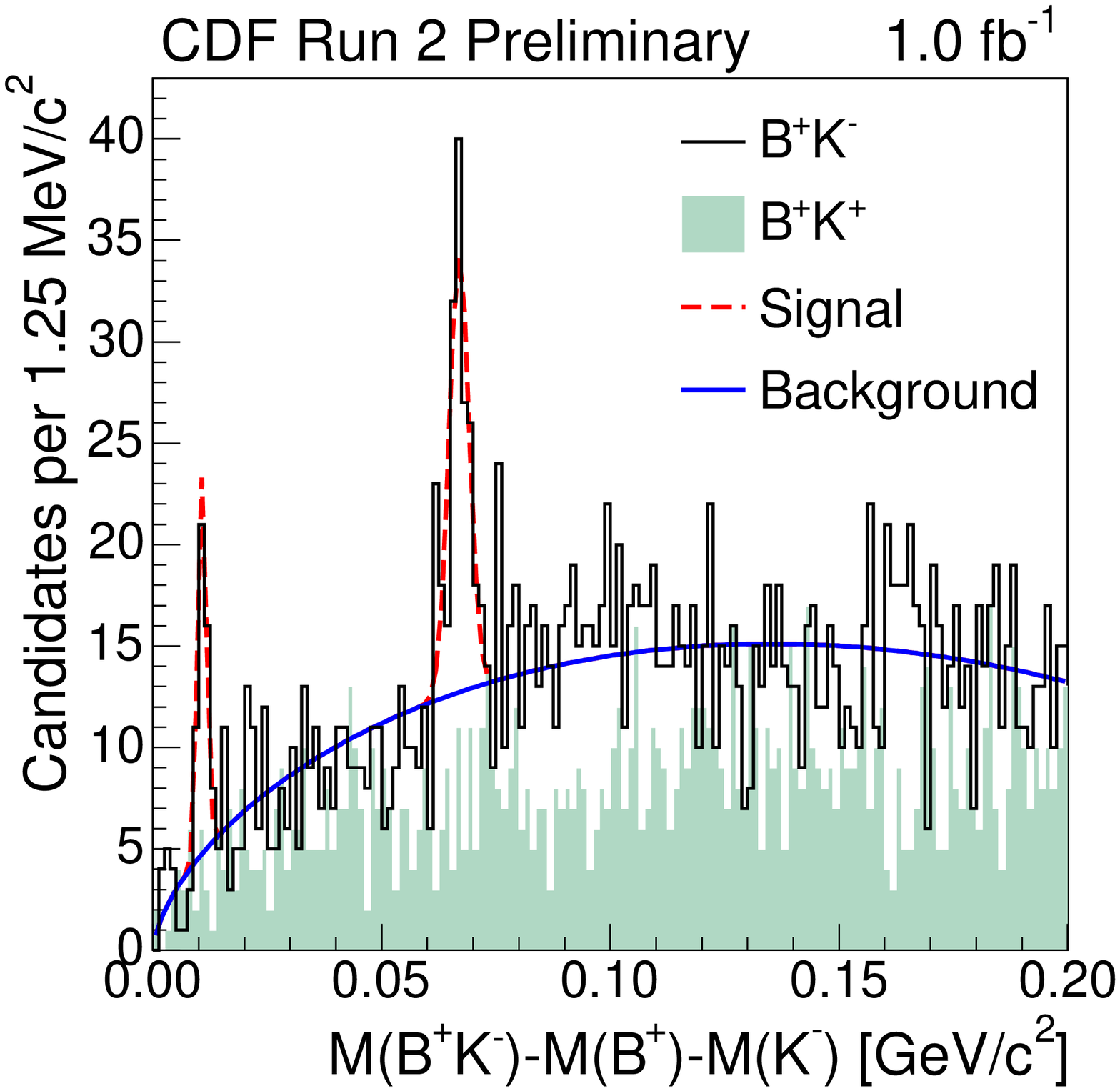}
  \includegraphics[width=0.56\textwidth]{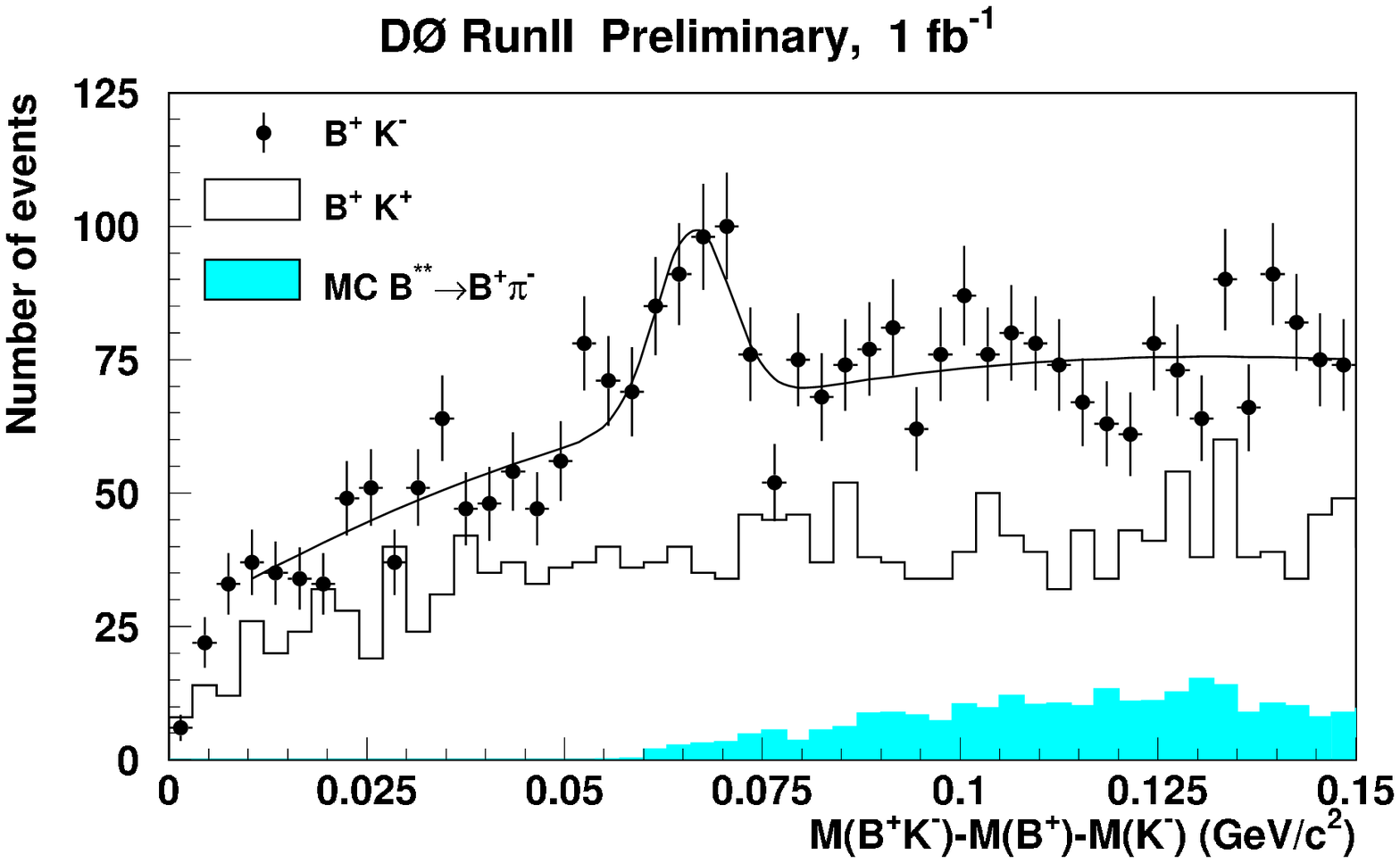}
  \caption{Distributions of the invariant mass difference $m(\mathrm{BK}) -
    m(\mathrm{B}) - m(\mathrm{K})$ for CDF (left) and D\O\ (right).}
  \label{fig:BK}
\end{figure}

Both experiments have also carried out analogous analyses of
$\mathrm{B_{s}^{\ast\ast 0}\rightarrow B^{+}K^{-}}$ decays. The
$\mathrm{B_{s}^{\ast\ast}}$ system is expected to replicate that of the
$\mathrm{B^{\ast\ast}}$. However, the decay $\mathrm{B_{s2}^{\ast 0}\rightarrow
  B^{\ast+}K^{-}}$ is expected to be suppressed strongly, so that at most two
peaks should be discernible. The results obtained are shown in
Fig.~\ref{fig:BK}.
Both experiments find a very clear peak around 67 MeV, which is attributed to
$\mathrm{B_{s2}^{\ast 0}\rightarrow B^{+}K^{-}}$ decays. In addition, CDF find
evidence for $\mathrm{B_{s1}^{0}\rightarrow B^{\ast +}K^{-}}$ decays, at a
$Q$-value of 11 MeV (D\O\ would not expect to observe a $\mathrm{B_{s1}^{0}}$
peak, given their observed mass splitting in the B system). The resulting mass
splitting of 10 MeV is fairly consistent with theoretical
predictions. Summarizing, both experiments obtain results consistent between B 
and $\mathrm{B_{s}}$; however, there is a lack of agreement between CDF and
D\O.

\subsection{$\Sigma_b$ baryons}
\label{sec:sigmab}

\begin{figure}[H]
  \begin{multicols}{2}
    CDF have used a sample of fully reconstructed
    $\Lambda_{\mathrm{b}}^{0}\rightarrow \Lambda_{\mathrm{c}}^{+}\pi^{-}$ decays to
    search for $\Sigma_{\mathrm{b}}$ baryons, in the decay
    $\Sigma_{\mathrm{b}}^{(\ast)\pm}\rightarrow \Lambda_{\mathrm{b}}^{0}\pi^{+}$. In
    these baryons, the light di-quark system has $I=1$ and $J^{P}=1^{+}$; adding the
    b quark leads to $J^{P}=\frac{1}{2}^{+}$ ($\Sigma_{\mathrm{b}}$) and 
    $J^{P}=\frac{3}{2}^{+}$ ($\Sigma^{\ast}_{\mathrm{b}}$). The
    $\Sigma_{\mathrm{b}}$--$\Lambda_{\mathrm{b}}$ mass difference is expected to be
    $\sim$ 200 MeV. The mass splitting between the systems is
    expected~\cite{rosner} to be
    $m(\Sigma^{\ast}_{\mathrm{b}})-m(\Sigma_{\mathrm{b}}) = 
    m_{\mathrm{c}}/m_{\mathrm{b}} \cdot
    (m(\Sigma^{\ast}_{\mathrm{c}})-m(\Sigma_{\mathrm{c}})) \approx 20$
    MeV; also a small splitting between states within the same
    isospin triplet is expected.

    In the analysis, the total decay widths are assumed to be saturated by single
    pion transitions, and constrained to the theory expectation of $\sim$ 8 MeV and
    15 MeV for the $\Sigma_{\mathrm{b}}$ and $\Sigma^{\ast}_{\mathrm{b}}$,
    respectively~\cite{koerner}. In addition, the isospin splittings are assumed to
    be the same for $\Sigma_{\mathrm{b}}$ and $\Sigma^{\ast}_{\mathrm{b}}$. The
    result is shown in Fig.~\ref{fig:Sigmab}, and is in good agreement with theory
    predictions: $m(\Sigma_{\mathrm{b}}^{-}) = 5816\pm 1\pm 1.7$ MeV,
    $m(\Sigma_{\mathrm{b}}^{+}) = 5808^{+2.3}_{-2.0} \pm 1.7$ MeV,
    $m(\Sigma_{\mathrm{b}}^{\ast -}) = 5837^{+2.1}_{-1.9} \pm 1.7$ MeV,
    $m(\Sigma_{\mathrm{b}}^{\ast +}) = 5829^{+1.6}_{-1.8} \pm 1.7$ MeV. (Note
    that not all of these results are independent.)

    \vspace*{2mm}
    \includegraphics[width=0.42\textwidth]{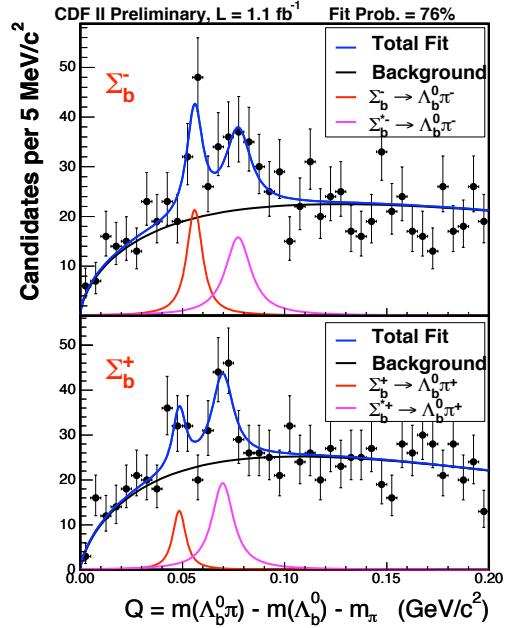}
    \caption{Distribution of the mass difference
      $m(\Lambda_{\mathrm{b}}\pi)-m(\Lambda_{\mathrm{b}})-m(\pi)$ for same-charge
      and opposite-charge $\Sigma_{\mathrm{b}}$ signals.}
    \label{fig:Sigmab}

  \end{multicols}
\end{figure}

\subsection{The $B_{c}$ meson}
\label{sec:Bc}

The $\mathrm{B_{c}}$ meson, the lowest mass bound state of a b and c quark,
has been observed already at LEP and in Run I of the Tevatron. However, only two
fully reconstructed event candidates were observed so far~\cite{opal}, in the
decay mode $\mathrm{B_{c}^{+}}\rightarrow J/\psi \pi^{+}$. The resulting
uncertainty on the mass was relatively large, about 60 MeV. CDF have now used
this same decay mode in their Run II data sample, requiring in addition that the
$J/\psi$ decay vertex be significantly displaced from the interaction
point. They observe a signal of $49.1 \pm 9.7$ events over a background of 34.1
events. The result of the mass fit is $m(\mathrm{B_{c}}) = 6276.5 \pm 4.0
(\mbox{stat.}) \pm 2.7 (\mbox{syst.})$ MeV. This represents an improvement of an
order of magnitude in accuracy over previous results.

\section{Lifetimes}
\label{sec:lifetimes}

Heavy Quark Effective Theory allows for a systematic expansion in orders of
$\alpha_{\mathrm{s}}$ and $1/m_{Q}$ of the total decay widths of heavy-quark
hadrons. As a result, precise predictions have been made for the ratios of
lifetimes of B hadrons. In the past, this led to the so-called
``$\Lambda_{\mathrm{b}}$ puzzle'', where the measured $\Lambda_{\mathrm{b}}$
lifetime ratio $\tau(\Lambda_{\mathrm{b}})/\tau(\mathrm{B_{d}})$ was
significantly below its theoretical expectation. Improved lattice 
gauge theory computations have since been made of B-hadron
lifetimes~\cite{tarantino}, and they have decreased this theoretical
expectation substantially to $0.88 \pm 0.05$, in fair agreement with
experiment.

Both CDF and D\O\ have measured $\tau(\Lambda_{\mathrm{b}})$ in the exclusive
decay $\Lambda_{\mathrm{b}}\rightarrow J/\psi
(\rightarrow\mu^{+}\mu^{-})\mathrm{\Lambda (\rightarrow p\pi)}$. This decay is
very similar to the decay 
$\mathrm{B_{d}}\rightarrow J/\psi \mathrm{K_{S} (\rightarrow\pi^{+}\pi^{-})}$,
allowing for a ``calibration'' of the
analyses using the precisely known $\mathrm{B_{d}}$ lifetime. The $J/\psi$ decay
vertex is combined with the reconstructed $\Lambda$ ($\mathrm{K_{S}}$) track to
yield the $\Lambda_{\mathrm{b}}$ ($\mathrm{B_{d}}$) vertex in the plane
perpendicular to the beam line; correcting for the boost yields the lifetime
estimate. Simultaneous unbinned maximum likelihood fits were made to the
invariant mass and lifetime distributions (and given that event-by-event
lifetime resolution estimates are used, the lifetime resolution distribution).

Both experiments measure a $\mathrm{B_{d}}$ lifetime compatible with the world
average. For the $\Lambda_{\mathrm{b}}$, D\O's measurement,
$\tau(\Lambda_{\mathrm{b}})/\tau(\mathrm{B_{d}}) =
0.811^{+0.096}_{-0.087}(\mbox{stat.}) \pm 0.034 (\mbox{syst.})$ is compatible 
with previous measurements and the theoretical predictions. However, the purer
and more precise CDF result, $\tau(\Lambda_{\mathrm{b}})/\tau(\mathrm{B_{d}}) =
1.018\pm 0.062 (\mbox{stat.})\pm 0.007 (\mbox{syst.})$, is much higher than
previous estimates.

D\O\ have measured the same quantity in the inclusive semileptonic decay
$\Lambda_{\mathrm{b}}\rightarrow\Lambda_{\mathrm{c}}\mu^{-}\bar{\nu}_{\mu}X$. This
measurement differs in many respects: it uses large statistics, but as it is an
\emph{inclusive} measurement a $\Lambda_{\mathrm{b}}$ peak is not observed.
Instead, the $\Lambda_{\mathrm{c}}$ is reconstructed in its decay mode
$\Lambda_{\mathrm{c}}\rightarrow\mathrm{K_{S} p}$. It is observed on top of a
large background, which is subtracted statistically, in bins of the visible
proper decay length. This quantity is corrected for the particles not
reconstructed in the $\Lambda_{\mathrm{b}}$ decay, the correction being modeled
by Monte Carlo simulations. The result, $\tau(\Lambda_{\mathrm{b}}) =
1.28^{+0.12}_{-0.11}(\mbox{stat.}) \pm 0.09 (\mbox{syst.})$ ps, is in good
agreement with D\O's measurement in the $J/\psi\Lambda$ channel. In conclusion,
the ``$\Lambda_{\mathrm{b}}$ puzzle'' cannot yet be considered resolved -- but
at least we know there is an \emph{experimental} issue to be addressed!

CDF have used their $J/\psi$ sample to measure also other lifetimes in
exclusively reconstructed decays. In particular, using the decays
$\mathrm{B^{+}\rightarrow J/\psi K^{+}}$ and $\mathrm{B_{s}^{0}\rightarrow
  J/\psi \phi}$, $\mathrm{\phi\rightarrow K^{+}K^{-}}$, they find
$\tau(\mathrm{B^{+}}) = 1.630 \pm 0.016 (\mbox{stat.})\pm 0.011 (\mbox{syst.})$
ps and
$\tau(\mathrm{B_{s}^{0}}) = 1.494 \pm 0.054 (\mbox{stat.})\pm 0.009
(\mbox{syst.})$ ps, respectively. These measurements are in good agreement with
earlier measurements, as well as with HQET predictions. Similarly, a new D\O\
measurement of the $\mathrm{B_{s}}$ lifetime using $\mathrm{B^{0}_{s}\rightarrow
  D_{s}^{-}\mu^{+}\nu_{\mu}}X$ decays, with the $\mathrm{D_{s}^{-}}$ identified
through $\mathrm{D_{s}^{-}}\rightarrow\phi\pi^{-}$,
$\phi\rightarrow\mathrm{K^{+}K^{-}}$, yields
$\tau(\mathrm{B_{s}^{0}}) = 1.398 \pm 0.044 (\mbox{stat.})^{+0.028}_{-0.025}
(\mbox{syst.})$ ps. This is in fair agreement with the current world average
result. It should be pointed out that both experiments' new $\mathrm{B_{s}}$
results have accuracies comparable to the present world average, offering good
hopes for more incisive tests of HQET.

Finally, CDF carried out a partial reconstruction of the decay
$\mathrm{B_{c}^{+}\rightarrow J/\psi(\rightarrow\mu^{+}\mu^{-}) e^{+}\nu_{e}}X$.
The $\mathrm{B_{c}}$ lifetime is expected to be much shorter,
$\tau(\mathrm{B_{c}}) = 0.48 \pm 0.05$ ps~\cite{kiselev}, than that of the
other weakly decaying B hadrons, because also the charm quark can decay. The
analysis attempts to isolate the exclusive decay $\mathrm{B_{c}^{+}}\rightarrow
J/\psi \mathrm{e^{+}\nu_{e}}$ using tight kinematic cuts. The challenge is a
proper understanding the instrumental backgrounds (many of which are
non-prompt), inferred from data, and from $\mathrm{b\bar{b}}$ production,
estimated using MC. The result, $\tau(\mathrm{B_{c}}) =
0.463^{+0.073}_{-0.065}(\mbox{stat.}) \pm 0.036(\mbox{syst.})$ ps, is in good
agreement with the present world average,
$\tau(\mathrm{B_{c}}) =
0.46^{+0.18}_{-0.16}(\mbox{stat.}) \pm 0.03(\mbox{syst.})$ ps. The new result is
the most precise one to date, and its accuracy approaches that of the
theoretical predictions.



\end{document}